\documentclass[aps,prl,showpacs,amsmath,amsfonts,amssymb,twocolumn]{revtex4} 
\usepackage{psfrag}
\usepackage[dvips]{graphicx}

\newcommand\RR{{\mathbb{R}}}
\newcommand\dd{{\mathrm{d}}}
\newcommand\ee{{\mathrm{e}}}

\DeclareMathOperator{\Order}{o}
\DeclareMathOperator{\Heaviside}{\Theta}
\DeclareMathOperator{\Dirac}{\delta}

\newtheorem{proposition}{Proposition}

\begin{document} 

FAU--TP3--07/1

\title{Phase Transitions from Saddles of the Potential Energy Landscape} 

\author{Michael Kastner} 
\email{Michael.Kastner@uni-bayreuth.de} 
\affiliation{Physikalisches Institut, Universit\"at Bayreuth, 95440 Bayreuth, Germany} 

\author{Oliver Schnetz} 
\affiliation{Institut f\"ur Theoretische Physik III, Friedrich-Alexander-Universit\"at Erlangen-N\"urnberg, Staudtstra{\ss}e 7, 91058 Erlangen, Germany} 

\author{Steffen Schreiber} 
\affiliation{Physikalisches Institut, Universit\"at Bayreuth, 95440 Bayreuth, Germany} 

\date{\today}
 
\begin{abstract}
The relation between saddle points of the potential of a classical many-particle system and the analyticity properties of its thermodynamic functions is studied. For finite systems, each saddle point is found to cause a nonanalyticity in the Boltzmann entropy, and the functional form of this nonanalytic term is derived. For large systems, the order of the nonanalytic term increases unboundedly, leading to an increasing differentiability of the entropy. Analyzing the contribution of the saddle points to the density of states in the thermodynamic limit, our results provide an explanation of how, and under which circumstances, saddle points of the potential energy landscape may (or may not) be at the origin of a phase transition in the thermodynamic limit. As an application, the puzzling observations by Risau-Gusman {\em et al}.\ on topological signatures of the spherical model are elucidated.
\end{abstract}

\pacs{05.70.Fh, 05.20.-y, 75.10.Hk} 

\maketitle 

Phase transitions, like the boiling and evaporating of water at a certain temperature and pressure, are common phenomena both in everyday life and in almost any branch of physics. Loosely speaking, a phase transition brings about a sudden change of the macroscopic properties of a many-particle system while smoothly varying a parameter (the temperature or the pressure in the above example). The mathematical description of phase transitions is conventionally based on (grand)canonical thermodynamic functions, relating their loss of analyticity to the occurrence of a phase transition. Such a nonanalytic behavior in a (grand)canonical thermodynamic function can occur only in the thermodynamic limit in which the number of degrees of freedom $N$ of the system goes to infinity \cite{Griffiths}.

Many researchers took it for granted that the same were true also for microcanonical thermodynamic functions. Recently, however, it was observed that the microcanonical entropy, or Boltzmann entropy, of a finite system is not necessarily analytic, 
and nonanalytic entropy functions of finite systems have been reported for certain classical models \cite{KaSchne:06,DunHil:06,CaKa:06}. 

In light of their conceptual similarity to the definition of a phase transition, it is tempting to regard finite-system nonanalyticities of the entropy as phase transitionlike phenomena. This point of view is advocated in \cite{DunHil:06}, and the authors of that reference argue that such nonanalyticities should also be measurable experimentally, at least in very small systems. However, such an interpretation is complicated by the fact that, as discussed in \cite{CaKa:06}, for typical models the number of nonanalytic points of the entropy increases unboundedly with the number of degrees of freedom $N$. 

Because of their typically large number, one might assume that nonanalyticities of the finite-system entropy were {\em unrelated}\/ to the occurrence of a phase transition in the thermodynamic limit. A theorem by Franzosi and Pettini \cite{FraPe:04}, however, indicates the existence of a relation between finite-system and infinite-system nonanalyticities (and we will come back to that theorem later). The purpose of the present Letter is to clarify and quantify the relation between saddle points of the potential energy and nonanalyticities of the entropy of finite systems and, in a second step, to use this relation in order to explain how, and under which conditions, a large number of ``weak'' nonanalyticities of the finite-system entropy may give rise to a single nonanalyticity (or a few) in the thermodynamic limit. At least in the common situation of ensemble equivalence, such a nonanalyticity of the infinite-system entropy will correspond to a nonanalyticity in the canonical free energy, and hence to a phase transition.

Our approach will be the following: We start from the case of classical many-particle systems, announcing an exact expression for the leading nonanalytic term that each saddle point of the potential energy landscape contributes to the density of states. This result, apart from its essential role for further steps of our analysis, is of interest in its own, giving a model-independent and quantitative account of nonanalyticities in the finite-system entropy as observed for the special cases in \cite{KaSchne:06,DunHil:06,CaKa:06}. Then, inspired by a calculation in \cite{FraPe}, the density of states is split into two terms: The first is a sum of the leading nonanalytic contributions stemming from all the saddle points, whereas the second contains the ``harmless'' rest. Performing the thermodynamic limit of the thus obtained expression, one can understand how, and under which conditions, the distribution of saddle points (as a function of the potential energy per particle) may cause a phase transition to take place. These results can be interpreted in terms of a recent approach relating phase transitions and configuration space topology, and we apply our findings to the spherical model, explaining the puzzling behavior of certain topological quantities reported in \cite{RiRiSta:05}.

{\em Preliminaries.}--- We consider classical systems of $N$ degrees of freedom, characterized by a Hamiltonian function of standard form,
\begin{equation}\label{eq:Hamiltonian}
H(p,q)=\frac{1}{2}\sum_{i=1}^N p_i^2 + V(q), 
\end{equation}
where $p=(p_1,\dotsc,p_N)$ is the vector of momenta and $q=(q_1,\dotsc,q_N)$ the vector of position coordinates. The restriction to a quadratic form in the momenta is not essential, but simplifies the following discussion. The potential $V$ is an analytic mapping from the configuration space $\Gamma_N\subseteq\RR^N$ onto the reals, and in general $V$ will have a number of {\em critical points}\/ (or {\em saddle points}) $q_\text{c}$ with vanishing exact differential, $\dd V(q_\text{c})=0$. We will assume in the following that all critical points of $V$ have a 
non-singular Hessian ${\mathfrak H}_V$. In this case, $V$ is called a {\em Morse function}, and one can argue that the restriction of the potential to this class of functions is not a serious limitation: Morse functions form a dense, open subset of all analytic functions, and hence are {\em generic}. 
Therefore, even if $V$ is not a Morse function, it can be made into one by adding an arbitrarily small perturbation \footnote{Alternatively, one could remove zero-eigenvalues of the Hessian arising from a symmetry of the system by fixing one or several position coordinates. Thermodynamic functions of such a ``reduced'' system will differ from the original ones only by a physically irrelevant constant; a detailed discussion of this issue will be published elsewhere.}.

{\em Finite system nonanalyticities.}--- Our aim is now to investigate the analyticity properties of the (configurational) density of states
\begin{equation}
\Omega_N(v)=\int_{\Gamma_N}\dd q \Dirac(V(q)-Nv)
\end{equation}
or, equivalently, of the entropy $s_N(v)=\ln[\Omega_N(v)]/N$ of a classical system of $N$ degrees of freedom, characterized by a Morse function $V$. The Morse property guarantees that all critical points of $V$ are isolated. We thus may reduce the discussion to the effect of a {\em single}\/ critical point of $V$ on the analyticity properties of the density of states $\Omega_N(v)$ as a function of the potential energy per degree of freedom $v$.
\begin{proposition}\label{prop:finite}
Let $V:G\to\RR$ be a Morse function with a single critical point $q_\text{c}$ in an open region $G\subset\RR^N$. Without loss of generality, we assume $V(q_\text{c})=0$. Then there exists a polynomial $P$ of degree less than $N/2$ such that at $v=0$ the density of states can be written in the form
\begin{equation}\label{eq:Omega_sep}
\Omega_N(v)=P(v)+\frac{h_{N,k}(v)}{\sqrt{\left|\det\left[{\mathfrak H}_{V}(q_\text{c})\right]\right|}}+\Order(v^{N/2-\epsilon})
\end{equation}
for any $\epsilon>0$, with the universal function
\begin{multline}\label{eq:h_Nk}
h_{N,k}(v)=\frac{(N\pi)^{N/2}}{\Gamma(\frac{N}{2})}\\
\times\begin{cases}
(-1)^{\frac{N-k}{2}} (-v)^{\frac{N-2}{2}} \Heaviside(-v) & \text{for $N,k$ odd},\\
(-1)^{\frac{k}{2}} \,v^{\frac{N-2}{2}} \Heaviside(v) & \text{for $k$ even},\\
(-1)^{\frac{k+1}{2}} \,v^{\frac{N-2}{2}}\,\pi^{-1}\ln|v| & \text{for $N$ even, $k$ odd}.\\
\end{cases}
\end{multline}
Here $k$ denotes the index of the critical point and $\Heaviside$ is the Heaviside step function.
\end{proposition}
This proposition gives a complete account of all types of nonanalyticities which can occur in $\Omega_N$ as the consequence of a {\em single}\/ critical point of a Morse function $V$. In the presence of several critical points, their nonanalytic contributions simply have to be added. For a proof of Proposition \ref{prop:finite}, the density of states $\Omega_N$ is calculated separately below and above the critical value $v=0$. By complex continuation it is possible to subtract both contributions and to evaluate the leading order of the difference \footnote{A detailed proof will be published elsewhere. A related, but weaker result has been announced in L.\ Spinelli, {\em Une approche topologique des transitions de phase}, PhD thesis, Universit\'e de Provence (1999).}.

The index $k$ of a critical point $q_\text{c}$ is the number of negative eigenvalues of the Hessian ${\mathfrak H}_V$ at $q_\text{c}$, and it determines whether a critical point is a minimum ($k=0$), a maximum ($k=N$), or a proper saddle ($k\in\{1,\dotsc,N-1\}$). Remarkably, the type of nonanalyticity described by $h_{N,k}$ in \eqref{eq:h_Nk} does not depend on the precise value of $k$, but only on whether $N$ and $k$ are odd or even. One can verify that in any of the three cases in \eqref{eq:h_Nk}, $\Omega_N$ is $\left\lfloor\frac{N-3}{2}\right\rfloor$-times continuously differentiable. This result is in agreement with the nonanalytic behavior of the exact solution for the density of states of the mean-field spherical model as reported in \cite{KaSchne:06}. In other words, the density of states $\Omega_N$ becomes ``smoother'' with increasing number of degrees of freedom, and already for moderate $N$ it supposedly will be impossible to observe such a finite-system nonanalyticity from noisy experimental or numerical data. At first sight one might therefore suspect that the nonanalyticities of the entropy are irrelevant for large systems and have no effect in the thermodynamic limit, but the following considerations will show that such an assertion is premature.

{\em Thermodynamic limit.}--- In a recent letter \cite{FraPe:04}, Franzosi and Pettini have discussed for a certain class of short-range models the relation between nonanalyticities of the entropy or the free energy in the thermodynamic limit and topology changes of the 
subsets
\begin{equation}
{\mathcal M}_v=\left\{q\in\Gamma_N\,\big|\,V(q)\leqslant Nv\right\}
\end{equation}
of configuration space $\Gamma_N$. Loosely speaking they found that, in order to have a phase transition at some potential energy $v_\text{t}$, a topology change of ${\mathcal M}_v$ is necessary to occur at $v=v_\text{t}$. 
Furthermore, in the Euler characteristic of ${\mathcal M}_v$ of several models analyzed, a signature was found at the phase transition energy $v_\text{t}$, corroborating the existence of a relation between phase transitions and topology changes \cite{CaPeCo:03,Angelani_etal:03}. 
It is a central proposition of Morse theory that, for potentials $V$ having the Morse property, each topology change of ${\mathcal M}_v$ at some value $v=v_\text{c}$ corresponds to one or several critical points $q_\text{c}$ of $V$ with critical value $v_\text{c}=V(q_\text{c})/N$ \cite{Hirsch}. Combining these results with Proposition \ref{prop:finite}, we infer that, despite their decreasing strength for large system sizes $N$, nonanalyticities of the finite-system entropy appear to be related to their infinite-system counterparts in some way.

This relation, however, is not one-to-one: critical points of $V$ are necessary, but by no means sufficient for a phase transition to occur. For several models studied, the number of critical levels $v_\text{c}$ was found to increase unboundedly with the number of degrees of freedom of the system, and the levels become dense on some interval in the thermodynamic limit \cite{CaPeCo:03,Angelani_etal:03,RiRiSta:05}. It is a pertinent open question to understand how, and under which conditions, nonanalyticities of the finite-system entropy may give rise to a phase transition in the thermodynamic limit. This issue which we will address in the following
, and---since a single critical point of $V$ is too ``weak'' to produce a phase transition in the thermodynamic limit---it is the distribution function of the very many, densely lying critical levels which will play important roles in this discussion.

By virtue of Proposition \ref{prop:finite}, we split the density of states of a system with $N$ degrees of freedom into two terms,
\begin{equation}\label{eq:splitting}
\Omega_N(v) = A_N(v) + B_N(v),
\end{equation}
 where $B_N$ contains the leading nonanalytic contributions from the critical points of $V$,
\begin{equation}\label{eq:B_N}
B_N(v) = \sum_{v_\text{c}}\sum_{q_\text{c}(v_\text{c})}\frac{h_{N,k}(v-v_\text{c})}{\sqrt{\big|\det\big[{\mathfrak H}_{V}(q_\text{c})\big]\big|}}.
\end{equation}
The sums extend over all critical values $v_\text{c}$ of $V$ and all critical values $q_\text{c}(v_\text{c})$ corresponding to the respective critical value. Due to the alternating signs of $h_{N,k}$ in \eqref{eq:h_Nk} we may consider $B_N$ as a {\em weighted}\/ alternating sum of the critical points. Then it is maybe not too surprising that the Euler characteristic of ${\mathcal M}_v$, being an {\em unweighted}\/ alternating sum of the critical points, was found to signal the presence of a phase transition for some models.

In order to perform the thermodynamic limit, we have to consider the entropy
\begin{equation}\label{eq:sv}
s(v)=\lim_{N\to\infty}\frac{1}{N}\ln\left[A_N(v)+B_N(v)\right],
\end{equation}
since, at least for a system with short-range interactions, we may expect this quantity to exist \cite{Ruelle}. It is important to note that $B_N$ is of order $\ee^N$ for $N\to\infty$ and therefore gives a finite contribution to $s$. Now we would like to deduce the (non)analyticity of the entropy $s$ from the properties of the thermodynamic limit expressions
\begin{equation}\label{eq:a_and_b}
a(v)=\lim_{N\to\infty}\frac{1}{N}\ln A_N(v),\quad b(v)=\lim_{N\to\infty}\frac{1}{N}\ln B_N(v).
\end{equation}
To this end it is instructive to rewrite Eq.\ \eqref{eq:sv} as
\begin{equation}\label{eq:sv2}
s(v)=\max\left\{a(v),b(v)\right\}.
\end{equation}
As a consequence of this expression, we cannot expect to draw any conclusions on $s$ from the knowledge of one single function $a$ or $b$, but only from the interplay of both. If, for example, we find a nonanalyticity in $b$, this nonanalytic behavior may be visible in $s$ in one case, but it may simply be overruled by a larger $a$ in another instance. Furthermore, a nonanalyticity in $s$ may arise from a {\em crossover}\/ between $a$ and $b$ when the dominant part in the maximum changes from $a$ to $b$ (or vice versa). 

A ``density splitting'' in \cite{FraPe}, similar to our Eq.~\eqref{eq:splitting}, indicates that one may expect the analytic part in $A_N$ to converge uniformly to an analytic function for well-behaved short-range potentials, whereas possible non-analytic contributions are dominated by $B_N$. Furthermore it appears plausible from the discussion in the preceding paragraphs that, at least for short-range models as covered by the theorem in \cite{FraPe:04}, nonanalyticities of the function $b$, stemming from the nonanalytic contributions of the critical points of the potential $V$, are crucial for the occurrence of a phase transition in the thermodynamic limit.

When performing the limit $N\to\infty$ we can for simplicity restrict the discussion to the subsequence with $N$ odd ($N$ even can be treated in a similar way). The sum over the critical values $v_\text{c}$ in \eqref{eq:B_N} is converted into an integral, yielding
\begin{widetext}
\begin{equation}\label{eq:B_N_limit}
B_N(v) \xrightarrow{N\to\infty} (2\pi\ee)^{N/2}\left[\int_{-\infty}^v\dd v_\text{c} \left[{\mathcal N}_0(v_\text{c})-{\mathcal N}_2(v_\text{c})\right](v-v_\text{c})^{(N-2)/2}+\int_v^{\infty}\dd v_\text{c} \left[{\mathcal N}_3(v_\text{c})-{\mathcal N}_1(v_\text{c})\right](v_\text{c}-v)^{(N-2)/2}\right],
\end{equation}
\end{widetext}
where the ${\mathcal N}_j$ are the limit distribution functions of the critical points 
with index $k=j\pmod 4$ where, as a consequence of \eqref{eq:B_N}, each point is weighted by $\left|\det[{\mathfrak H}_V(q_\text{c})]\right|^{-1/2}$. Note that $B_N$ depends on the {\em differences}\/ between the densities of certain types of critical points, and to illustrate the implications of this fact, we will now apply Eq.\ \eqref{eq:B_N_limit} to the spherical model. This will allow us to understand the puzzling observations made by Risau-Gusman {\em et al.}\ \cite{RiRiSta:05} when studying the topology of the subsets ${\mathcal M}_v$ of this model.

{\em Application to the spherical model.}--- This model was introduced by Berlin and Kac \cite{BerKac:52} as an exactly solvable caricature of the Ising model of a ferromagnet. Its configuration space $\Gamma_N$ is an $(N-1)$-sphere with radius $\sqrt{N}$, and the potential reads
\begin{equation}\label{eq:V_spherical}
V:\Gamma_N\to\RR,\qquad q\mapsto -\sum_{i,j=1}^N J_{ij}q_i q_j.
\end{equation}
The degrees of freedom $q_i$ of the model are placed on a $d$-dimensional cubic lattice, and we consider a coupling matrix $J$ with elements $J_{ij}=\frac{1}{2}$ when $i$ and $j$ are neighboring sites on the lattice, and $J_{ij}=0$ otherwise. The spherical model is solvable in the thermodynamic limit for arbitrary $d$, and a phase transition from a ferromagnetic phase at low temperatures to a paramagnetic phase at high temperatures (or energies) occurs for all $d\geqslant3$. 

In a recent letter \cite{RiRiSta:05}, Risau-Gusman {\em et al.}\ reported a study of the topology of the subsets ${\mathcal M}_v\subseteq\Gamma_N$ of this model. Computing the deformation retract of the ${\mathcal M}_v$, they found that their topology is determined by the spectral density of the coupling matrix $J$. In the limit $N\to\infty$, this spectral density is given by
\begin{equation}\label{eq:cv}
c(v)=\frac{1}{\pi}\int_0^\infty \dd x \cos(xv)[J_0(x)]^d,
\end{equation}
where $J_0$ is the Bessel function of order zero. Depending on the dimension $d$, the function $c$ has one or several nonanalytic points, however---inconsistent with an expected relation between topology changes and phase transitions---none of these points coincides with the transition energy $v_\text{t}$ of the spherical model.

Instead of following the approach in \cite{RiRiSta:05}, we analyze the subsets ${\mathcal M}_v$ by means of Morse theory. Although the potential $V$ of the spherical model in \eqref{eq:V_spherical} is not a Morse function, it can be made into one by adding a small coupling to a magnetic field $h$. In the simultaneous limit $h\to0$ and $N\to\infty$ one can relate the critical densities ${\mathcal N}_j$ in \eqref{eq:B_N_limit} to the limit distribution $c$, yielding ${\mathcal N}_0={\mathcal N}_1={\mathcal N}_2={\mathcal N}_3$. We thus find $b=0$ for the spherical model, which implies that the contribution of the critical points of $V$ to the entropy vanishes in the thermodynamic limit. This explains why no effect of the nonanalytic points of $c$ is visible in the thermodynamic functions of the spherical model.

At first sight this finding seems in conflict with the previous discussion where the relevance of the $b$-term was emphasized. This apparent contradiction can be understood by noting that the spherical model is not genuinely short-range. Although the interaction term in \eqref{eq:V_spherical} is restricted to pairs of nearest neighbors, the spherical configuration space renders the interaction effectively long-range. For long-range systems however, as discussed in \cite{HaKa:05}, the connection between phase transitions and configuration space topology is not valid in general and we cannot assume the function $a$ in \eqref{eq:sv2} to be analytic. Similar to the observations made in \cite{HaKa:05} for a different long-range model, we suspect the entropy $s(v,m)$ as a function of the potential energy $v$ and the magnetization $m$ to be a nonconcave function, such that the nonanalyticity in $s(v)=\max_m s(v,m)$ arises from the maximization over $m$.

{\em Summary.}--- We have analyzed the relation between saddle points of the potential energy $V$ of classical $N$-particle systems and the analyticity properties of thermodynamic functions. For finite systems, each saddle point $q_\text{c}$ was found to cause a nonanalyticity in the entropy $s_N(v)$ at the value $v=V(q_\text{c})/N$ of the potential energy, and the functional form of the nonanalytic term is specified in Proposition~\ref{prop:finite}. Since the number of saddle points is expected to grow exponentially with $N$ for generic potentials, we arrive at the remarkable conclusion that the finite-system entropy may have a large number of nonanalytic points.
Considering the contribution of very many saddle points becoming dense in the thermodynamic limit, we discussed how, despite the increasing differentiability of $s_N$, a continuous distribution of saddle points may lead to a nonanalyticity in the infinite-system entropy. Interpreting our findings in the spirit of the topological approach to phase transition, our results indicate under which conditions topology changes of the subsets ${\mathcal M}_v$ may lead to a phase transition in the thermodynamic limit. An application of these findings to the spherical model allowed us to understand the puzzling observations of Risau-Gusman {\em et al.}\ that, for this model, topological signatures and phase transition energy do not coincide.

Conceptually, our study explores the connection between phase transition theory and the study of energy landscapes, a rapidly developing field with applications, among others, to clusters, biomolecules, and glassy systems \cite{Wales}. One might hope to profit from the considerable knowledge on the relation between energy landscapes and dynamical properties in future work.


\bibliographystyle{h-physrev}
\bibliography{PTsaddles.bib}

\end{document}